\documentclass[12pt,a4wide,3p,authoryear,nopreprintline]{article}
\usepackage{natbib}
\usepackage[]{geometry}
\usepackage{a4wide}
\usepackage{enumerate}
\usepackage{float}
\usepackage{caption}
\usepackage[bookmarks]{hyperref}

\usepackage{parskip}

\usepackage{color}
\usepackage{graphicx}
\usepackage{booktabs}
\usepackage{capt-of}
\usepackage{floatpag}
\usepackage[utf8x]{inputenc}  
\usepackage{amsmath,amsfonts,amssymb,amsthm}
\usepackage{mathtools} 
\usepackage{dsfont}
\usepackage{rotating}
\usepackage{multicol}
\newtheorem{mydef}{Definition}

\usepackage[font=small,labelfont=bf]{caption}

\allowdisplaybreaks

\begin{document} 
\vspace*{3em}
\begin{center}
\begin{LARGE}
Practical Relevance: A Formal Definition
\end{LARGE}\\[1.5em]
\begin{large}
Patrick Schwaferts \quad \quad Thomas Augustin
\end{large}\\[1em]
\begin{scriptsize}
patrick.schwaferts@stat.uni-muenchen.de\\
thomas.augustin@stat.uni-muenchen.de\\[1em]
Ludwig-Maximilians-Universität Munich\\
Department of Statistics\\
Methodological Foundations of Statistics and its Applications\\
Ludwigsstraße 33, 80539 Munich, Germany\\[5em]
\end{scriptsize}
\end{center}

\begin{abstract}
There is a general agreement that it is important to consider the practical relevance of an effect in addition to its statistical significance, yet a formal definition of practical relevance is still pending and shall be provided within this paper.
It appears that an underlying decision problem, characterized by actions and a loss function, is required to define the notion of practical relevance, rendering it a decision theoretic concept.
In the context of hypothesis-based analyses, the notion of practical relevance relates to specifying the hypotheses reasonably, such that the null hypothesis does not contain only a single parameter null value, but also all parameter values that are equivalent to the null value on a practical level. In that regard, the definition of practical relevance is also extended into the context of hypotheses. The formal elaborations on the notion of practical relevance within this paper indicate that, typically, a specific decision problem is implicitly assumed when dealing with the practical relevance of an effect or some results. As a consequence, involving decision theoretic considerations into a statistical analysis suggests itself by the mere nature of the notion of practical relevance.\\

Keywords: Practical Significance, Practical Relevance, Nil Hypothesis, Decision Theory, Reproducibility Crisis, Null Hypothesis Significance Testing
\end{abstract}

\section{Introduction}\label{sec:introduction}

More than twenty years have passed since \citet{Kirk1996} urged to consider the practical relevance of research results in addition to their statistical significance with his paper entitled ``Practical Significance: A Concept Whose Time Has Come,'' yet no formal definition of this concept is currently available. Instead, merely stating that effect sizes are ``measures'' \citep[see e.g.][]{Ellis2003} or ``indices'' \citep[see e.g.][]{Thompson2002, Hojat2004} of practical significance which indicate if results are ``meaningful'' \citep[see e.g.][]{Vaske2002} or ``useful'' \citep[see e.g.][]{Kirk1996} seemed to be sufficient. This, however, is by no means a proper mathematical incorporation of the notion of practical relevance (or practical significance) within the frameworks of statistical methodologies. In that, this paper attempts to provide a formal definition of practical relevance.

There are two different lines of research that lead to a definition of practical relevance, which appear to be closely related. The first one is mentioned above and directly concerned with the practical relevance of an effect \citep[see e.g.][]{Kirk1996}. The second one is based on the criticism \citep[see e.g.][]{Cohen1994} that null hypotheses within the omnipresent approach of null hypothesis significance testing (NHST) typically hypothesize a single parameter value representing a zero effect size, yet this is not of interest as it does not matter if the effect is literally zero, but only practically zero (e.g.\ a parameter value of, say, $0.01$ is not exactly zero but might be practically equivalent to zero in most cases). In that regard, currently promoted statistical methods use reasonably specified null hypotheses, considering smallest effect sizes of interest in equivalence tests \citep[see e.g.][]{Lakens2017,Lakens2018}, regions of practical equivalence (ROPE) around the null value in Bayesian decision rules \citep[see e.g.][]{Kruschke2015,Kruschke2018}, or interval-valued null hypotheses in the context of Bayes factors \citep[see e.g.][]{Morey2011, Hoijtink2019,Heck2020}.

By evaluating both of these lines of research, it seems that practical relevance can only be described by referring to a (potentially implicitly assumed) decision problem in which one of two actions should be chosen, one being associated with an effect that is practically zero and one being associated with an effect that is practically relevant (i.e.\ non-zero). In that regard, the context of decision making is necessary to formalize this situation and to provide a definition of practical relevance. In accordance with both of these lines of research, two different definitions of practical relevance might be distinguished, one referring to effects (Section~\ref{sec:effects}) and one referring to hypotheses (Section~\ref{sec:hypotheses}). The implications of these definitions for applied sciences will be discussed (Section~\ref{sec:practice}).

\section{Practical Relevance of an Effect}\label{sec:effects}

\subsection{Context}\label{sec:effect_context}

Practical significance is typically introduced in research papers by stating that effect sizes are measures of it which indicate the importance of the result. However, there is a general agreement that effect sizes are not synonymous with practical significance \citep[see e.g.][]{Vaske2002, Peeters2016} and that there is more to practical significance than the mere size of the effect. In this regard, \citet[p.~213, line breaks added]{Kirk2001} states that ``[r]esearchers want to answer three basic questions:
\begin{itemize}
\item[(a)] Is an observed effect real or should it be attributed to chance?
\item[(b)] If the effect is real, how large is it? and
\item[(c)] Is the effect large enough to be useful?''
\end{itemize}

The first question (a) might be answered by assessing the uncertainty within the observed effect, which is conventionally done by conducting a statistical test, although recently different approaches are promoted, such as Bayesian statistics \citep[see e.g.][]{Schoot2017} or estimation methods that acknowledge the available uncertainty, such as confidence intervals \citep[see e.g.][]{Cumming2014}.
The second question (b) might be answered by the effect size estimate, however, the answer to the third question (c) is more difficult. The usefulness of an effect naturally depends on what it is used for.

Consider the following stereotypical example that is aptly illustrated by \citet[p.~279]{Sullivan2012}:

\begin{quote}
A commonly cited example of this problem is the Physicians Health Study of aspirin to prevent myocardial infarction (MI). [\citep{Bartolucci2011}] In more than $22000$ subjects over an average of 5 years, aspirin was associated with a reduction in MI (although not in overall cardiovascular mortality) that was highly statistically significant: $P < .00001$. The study was terminated early due to the conclusive evidence, and aspirin was recommended for general prevention. However, the effect size was very small: a risk difference of $0.77\%$ with $r^2 = .001$ -- an extremely small effect size. As a result of that study, many people were advised to take aspirin who would not experience benefit yet were also at risk for adverse effects. Further studies found even smaller effects, and the recommendation to use aspirin has since been modified.
\end{quote}

Similar examples can be found easily, e.g.\ analogously about whether a medication should be administered or not in the context of a certain disease \citep{Baicus2009}, about the gain in knowledge and the ability to think critically of university students in the context of deciding about different teaching methods \citep{Peeters2016}, or fictitiously about assessing IQ differences between two arbitrary groups of students that might lead to decisions about where to erect new schools for talented students \citep{Thompson1993}.

All those examples have two characteristics in common.
First, a decision has to be be guided (e.g.\ administer aspirin to prevent MI or not), such that the usefulness of the reported effect can be assessed w.r.t.\ this decision, creating the framework to answer question~(c).
Second, there is agreement on the lack of practical relevance of the reported effect, such that it seems reasonable to decide as if no effect was present. Accordingly, a way to determine how to decide for each possible effect is implicitly employed, which allows to implicitly answer question (c) in the framework of the decision of interest.

Both a decision and a way to decide for each possible effect are central components of statistical decision theory \citep[see e.g.][]{Berger1985, Robert2007}. Therefore, it suggests itself to employ decision theoretic concepts in order to define the practical relevance of an effect.

\subsection{Formal Definition}\label{sec:effect_definition}

Assume the observed data $y$ are modelled as realization of the random variable $Y$ with parametric density $f(y| \theta , \varphi)$, where $\theta$ is the effect parameter of interest (e.g.\ a standardized or non-standardized difference in means between two groups or a correlation between two features) and $\varphi$ is a vector of nuisance parameters being not of interest. Without loss of generality, the effect parameter $\theta$ is such that $\theta = 0$ indicates the absence of an effect (else available parameters might be transformed accordingly).

A decision should be guided between two actions $a_0$ and $a_1$, where $a_0$ should denote the action that is appropriate if the effect is absent, i.e.\ if $\theta = 0$.

For each possible effect $\theta \in \Theta$ within the parameter space $\Theta$, each of both actions has certain consequences and the ``badness'' of these consequences is quantified by a loss function
\begin{equation}
L: \Theta \times \mathcal{A} \to \mathbb{R}^+_0 :
(\theta , a) \mapsto L(\theta,a) \, ,
\end{equation}
where $\mathcal{A} = \lbrace a_0 , a_1 \rbrace$ is the action space. So, e.g.\ $L(0,a_0)$ and $L(0,a_1)$ quantify how bad it would be to decide for $a_0$ and $a_1$, respectively, if $\theta = 0$ is true.

The loss function might be seen as representing the consequences of the applied decision on a formal level. In order to obtain such a loss function for a certain decision problem, one needs to think about the consequences of deciding for $a_0$ or $a_1$ in dependence of the parameter value $\theta$ and then find a function $L$ that mathematically represents the badness of those consequences. Naturally, information about these consequences might be vague, such that specifying a loss function unambiguously might be difficult, as many different loss functions might be in accordance with the available vague information. However, the fact that it might be difficult to specify such a loss function unambiguously does not prohibit that an appropriate mathematical formulation of the notion of practical relevance is embedded within a decision theoretic context. The aim of this paper is to derive the concept of practical relevance on a formal level to gain a better mathematical understanding of it. Therefore, assume for now that such a loss function is available.

For each effect $\theta$ the action with smaller loss should be preferred. Therefore, one of the following holds for each effect $\theta$:
\begin{itemize}
\item If $L(a_0, \theta) < L(a_1,\theta)$, then $a_0$ is preferred over $a_1$.
\item If $L(a_1, \theta) < L(a_0,\theta)$, then $a_1$ is preferred over $a_0$.
\item If $L(a_0, \theta) = L(a_1,\theta)$, then there is no preference between $a_0$ and $a_1$.
\end{itemize}

Due to choosing $a_0$ to be appropriate in the absence of an effect, $L(0, a_0)$ is smaller than $L(0, a_1)$. Intuitively, other effects $\theta$ that also prefer $a_0$ over $a_1$ cannot be practically relevant as they lead to the same decision as the absence of an effect. Consequently, it must be those effects $\theta$ that prefer $a_1$ over $a_0$ which are practically relevant.

Summing up, these considerations lead to the definition of the practical relevance of an effect:

\begin{mydef}[Practical Relevance of an Effect]
Within this framework, an effect $\theta$ is practically relevant (or practically significant) w.r.t.\ the actions $a_0$, $a_1$, and the corresponding loss function $L$, if $a_1$ is preferred over $a_0$, i.e.\ if
\begin{equation}
L(\theta , a_1) < L(\theta , a_0) \, ,
\end{equation}
else the effect $\theta$ is negligible (or practically zero) w.r.t.\ these actions and this loss function.
\end{mydef}

The terms practical relevance and practical significance shall be used interchangeably, because although these considerations arise from the frequentist literature about the practical significance of an effect, they also apply to the Bayesian context in which the term ``significance'' is typically avoided.

\subsection{Example}\label{sec:effect_example}

An artificial coin flipping example shall be used as illustration. Person A offers a gamble to person B: Person A will flip a presumably fair coin 10 times. If the number of heads is between 4 and 6, then person B wins, else person A wins.
However, person B manages to check the coin in advance. To do so, person B plans to flip the coin several times to estimate its probability of heads $\pi$ and calculate the coins' bias $b:= \pi - 0.5 \in [-0.5,0.5] \, ,$
which represents the effect parameter.
Depending on the outcome, person B might think about accusing person A of cheating. In that, the possible actions of person B are:
\begin{itemize}
\item[$a_0$:] not accuse person A of cheating
\item[$a_1$:] accuse person A of cheating
\end{itemize}

\begin{figure}
	\centering
  \includegraphics[width=0.8\textwidth]{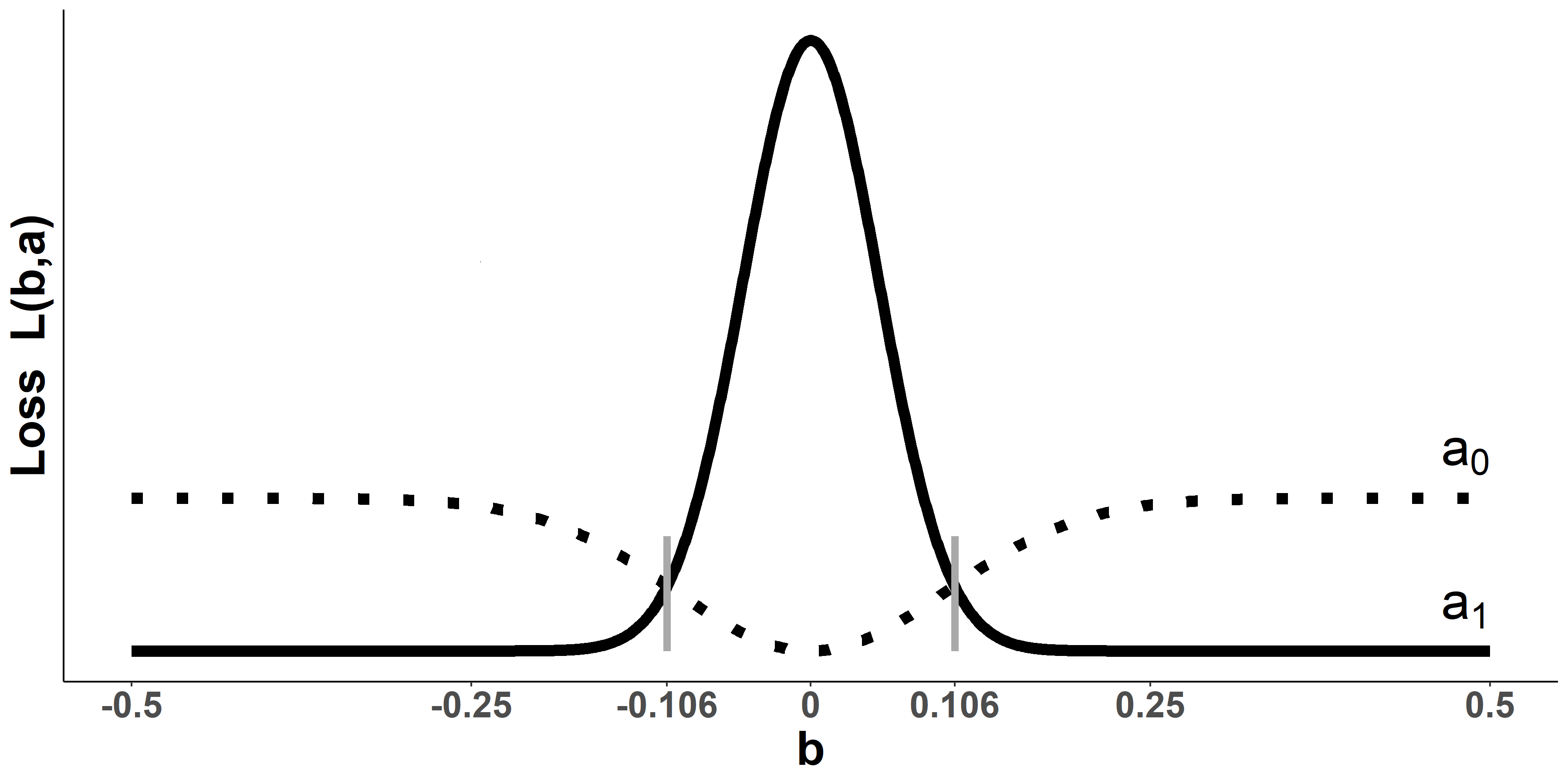}
	\caption{Example: Loss Function. The loss value ($y$-axis) is depicted for each possible bias~$b$ ($x$-axis) and each of both actions $a_0$ (dotted line) and $a_1$ (solid line). Both lines cross at bias values $b=-0.106$ and $b=0.106$. Within this example, this loss function was arbitrarily chosen and is treated as given.}
	\label{fig:loss_function}
\end{figure}

For each potential bias $b$, person B assesses how bad each of both actions would be, and reasons that this badness might be represented mathematically by the loss function depicted in Figure~\ref{fig:loss_function} (As mentioned, specifying such a loss function unambiguously in an applied context is a difficult task. Therefore, for this example, this loss function shall be treated as given.).

With this loss function, it appears that bias values $b$ within $B_0 = [ -0.106 ,0.106 ]$ are negligible and bias values within $B_1 = [-0.5 ,-0.106 ) \cup ( 0.106 , 0.5]$ are practically relevant w.r.t.\ to the decision of person B.

\subsection{Discussion}

As made explicit by this definition, the practical relevance or negligibility of an effect involves certain actions and a specific loss function. For different actions or with a different loss function, different effects might be practically relevant or negligible. In that, it is recommended to explicitly state the actions and describe corresponding consequences in an applied context.

In order to keep the formal definitions simple, an effect $\theta$ for which both actions have the same loss, i.e.\ $L(a_0, \theta) = L(a_1,\theta)$, is arbitrarily treated as negligible. Nevertheless, it might be equally valid to treat it as practically relevant, yet this will hardly be of importance in applied investigations.

The decision theoretic concepts employed within this definition are the actions themselves and the loss function. As latter allows to determine which action should be preferred for each effect $\theta$, these concepts are exactly those needed to answer question (c) about whether the effect $\theta$ is useful (i.e.\ practically relevant) or not \citep[][]{Kirk2001}. Naturally, the observed data $y$ have not been involved so far, as determining which potential effects $\theta$ are practically relevant is possible (and actually necessary) before collecting the data.

Although only few decision theoretic concepts (actions and loss function) are necessary to define the practical relevance (or negligibility) of an effect, decision theory is an extensive framework for deciding in the face of uncertainty. In addition to well-founded work on how to include data in this decision process, it is possible to include a variety of different external information, e.g.\ within the loss function, within a prior distribution or by choosing a certain decision paradigm or decision principle \citep[see e.g.][]{Berger1985}.

The reliance on decision theoretic concepts has already been anticipated by those scientists dealing with practical significance. For example, \citet[p.~103]{Pintea2010} emphasized the importance of the decision for which a research result is used: ``An increasing number of authors underline the gap between researchers who only report the statistical significance of their results and practitioners who need relevant information for their decisions in clinical, counseling, educational, and organizational practice.''
Also the necessity to include the usefulness or value of each effect into a statistical analysis, which can be achieved by a loss function, was highlighted e.g.\ by \citet[p.~365]{Thompson1993}: ``If the computer package did not ask you your values prior to its analysis, it could not have considered your value system in calculating $p$[-value]s, and so $p$[-value]s cannot be blithely used to infer the value of research results.''

\section{Practical Relevance in the Context of Hypotheses}\label{sec:hypotheses}

\subsection{Context}\label{sec:hypo_context}

An additional characteristic of the examples, that lead to the definition of a practically relevant effect, is that a conventional null hypothesis significance test leads to a questionable result.
Null hypothesis significance testing (NHST) typically involves a null hypothesis that hypothesizes only a single parameter value representing a zero effect -- such that this null hypothesis is frequently referred to as nil hypothesis \citep[see e.g.][]{Cohen1994} -- and a general alternative hypothesis that hypothesizes all other possible parameter values.
For over 80 years, this approach has been subject to the critique of being not of interest \citep[see e.g.][]{Berkson1938, Cohen1994, Gigerenzer2004}.
In that, the conventional NHST approach might lead to a conclusion which favors the alternative hypothesis (commonly interpreted as presence of an effect) even if the observed effect is negligible (as e.g.\ in the aspirin example). This is because also negligible effects are hypothesized within such a general alternative hypothesis.

Similar issues might arise in the context of Bayes factors \citep[][]{Jeffreys1961, Kass1995, Goenen2005, Rouder2009}, a Bayesian alternative to frequentist hypothesis tests. Frequently, Bayes factors are also calculated with sharp null hypotheses \citep[see e.g.][]{Jeffreys1961,Rouder2009,Rouder2018b, Rouder2018, Lakens2020}, such that corresponding alternative hypotheses might also contain negligible effects. However, it has to be noted that there are exceptions, which consider interval-valued null hypotheses \citep[see e.g.][]{Morey2011, Hoijtink2019,Heck2020}.

Critics claim that -- using the terminology of this paper -- the null hypothesis should hypothesize all negligible effects, not only the zero effect, and that the alternative hypothesis should hypothesize practically relevant effects \citep[see e.g.][]{Berger1985, Morey2011, Lakens2018, Kruschke2018, Blume2019}.
As the definition of a practically relevant or a negligible effect requires the presence of two actions and a loss function, so does meeting this claim. In that, appropriately specified hypotheses are to be defined within the context of decision theory.

\subsection{Formal Definition}\label{sec:hypo_definition}

Continuing with the previous notation, hypotheses about the effect $\theta$ are subsets $\Theta_0$ and $\Theta_1$ of the parameter space $\Theta$:
\begin{equation}\label{eq:hypotheses}
H_0: \theta \in \Theta_0 \quad \textrm{vs.} \quad H_1: \theta \in \Theta_1 \, .
\end{equation}
In NHST, $\Theta_0 = \lbrace 0 \rbrace$ contains only the zero effect and $\Theta_1 = \lbrace \theta \in \Theta \vert \theta \neq 0 \rbrace$ contains all other effect values.
However, for a given decision between the actions $a_0$ and $a_1$ and a given loss function $L$, this $\Theta_1$ (representing the general alternative) typically contains also effects~$\theta$ that are negligible w.r.t. the given decision problem, resulting in the critique outlined above.

Instead, $\Theta_0$ should contain no practically relevant effects and $\Theta_1$ should contain no negligible effects, yet this is possible in two different ways:
\begin{itemize}
\item All effects within the subsets $\Theta_0$ and $\Theta_1$ are negligible and practically relevant, respectively.
\item All negligible and practically relevant effects are within the subsets $\Theta_0$ and $\Theta_1$, respectively.
\end{itemize}

In the first case, there might still be (negligible or practically relevant) effects left which are not contained in either hypothesis, i.e.\ $\Theta = \Theta_0 \cup \Theta_1$ need not hold, and in the second case, any effect is contained in one of the hypotheses, i.e.\ $\Theta = \Theta_0 \cup \Theta_1$ holds. Accordingly, in the former case the hypotheses might incorporate the notion of practical relevance only partially, in the latter case even completely, leading to the following definition\footnote{Within this definition, hypotheses are separated in a mathematically exact way based on the loss function: Even for very small differences between the loss values of both actions, an effect value is placed within one of the hypotheses if its loss value is smaller than the other loss value. In this regard, an idealized precise underlying loss function is assumed and dealt with in an numerical exact way, leading to a precise boundary between both hypotheses.
In contrast, a more applied and less idealized view of the loss function might be to allow a rather vague boundary between the hypotheses, consisting of a range of different effect values. Consequently, all these effects, which characterize this vague boundary, cannot differentiate between both hypotheses and might be (arbitrarily) referred to negligible. With regard to its interpretation, this emphasizes a rather subjective nature of the loss function: The badness of the consequences of the different actions are \textit{perceived} as somehow equivalent for these ranges of effects.}:

\begin{mydef}[Practical Relevance w.r.t.\ Hypotheses]
Within this framework, two hypotheses about an effect $\theta$ (equation \ref{eq:hypotheses}) \textbf{completely} incorporate the notion of practical relevance (or practical significance) w.r.t.\ two associated actions $a_0$, $a_1$ and the corresponding loss function $L$, if $\Theta_1$ contains \textbf{all} practically relevant effects and $\Theta_0$ contains \textbf{all} negligible effects, i.e.\
\begin{align}
&\forall \theta \in \Theta: L(\theta , a_0) \leq L(\theta , a_1) \Rightarrow \theta \in \Theta_0\\
&\forall \theta \in \Theta: L(\theta , a_1) < L(\theta , a_0) \Rightarrow \theta \in \Theta_1 \, .
\end{align}
These hypotheses (equation \ref{eq:hypotheses}) \textbf{partially} incorporate the notion of practical relevance (or practical significance) w.r.t.\ these actions and this loss function, if $\Theta_1$ contains \textbf{only} practically relevant effects and $\Theta_0$ contains \textbf{only} negligible effects, i.e.\
\begin{align}
&\forall \theta \in \Theta_0: L(\theta , a_0) \leq L(\theta , a_1)\\
&\forall \theta \in \Theta_1: L(\theta , a_1) < L(\theta , a_0) \, .
\end{align}
\end{mydef}

\subsection{Discussion and Example}

Naturally, if hypotheses incorporate the notion of practical relevance completely they also do so partially. Continuing the previous coin flipping example, the hypotheses
\begin{align}
&H_0: b \in [ -0.106 ,0.106 ] \quad \textrm{vs.}\nonumber\\
&H_1: b \in [-0.5 ,-0.106 ) \cup ( 0.106 , 0.5] \label{eq:exp_hypo_complete}
\end{align}
incorporate the notion of practical relevance both completely and partially w.r.t.\ the underlying actions and loss function.

However, this implication does not hold in the reverse direction. For example, the (arbitrarily chosen) hypotheses
\begin{equation}\label{eq:exp_hypo_partial}
H_0: b \in \{0\} \quad \textrm{vs.} \quad H_1: b \in \{0.3\}
\end{equation}
incorporate the notion of practical relevance only partially, but not completely, w.r.t.\ the underlying actions and loss function.

Incorporating the notion of practical relevance into hypotheses only partially, and not completely, is equivalent to a restriction on the parameter space. The union of both of these hypotheses constitutes a new restricted parameter space $\{0,0.3\}$, in which they (equation~(\ref{eq:exp_hypo_partial})) incorporate the notion of practical relevance completely. Accordingly, this implies that all other parameter values $[-0.5,0.5] \setminus \{0,0.3\}$ are irrelevant within the conducted analysis. This is a strong claim and needs to be justified. Therefore, it is recommended to primarily employ a parameter space that is meaningful in the context of the sampling distribution and then derive hypotheses that incorporate the notion of practical relevance completely w.r.t.\ this parameter space and the underlying decision problem.

\section{In Practice} \label{sec:practice}

This paper offers a formal definition of the concept of practical relevance (or practical significance) for both effects and hypotheses.
It appears that a proper definition of this concept depends on an underlying decision problem. Without such a decision problem, it is neither possible to assess the practical relevance of an effect nor to specify practically relevant hypotheses, as their practical relevance naturally depends on what they are used for.

The main goal of this elaboration is to understand the notion of practical relevance on a formal level. This is necessary to include it into a statistical analysis in a mathematically correct manner. Without it, the discussion about the practical relevance of an observed effect or of some research results is of mere qualitative nature. The researcher interprets the observed effect and the results, and integrates them into the broader research context. If this context relates to a practical problem, there will be a judgment about the practical relevance of the results. Yet, this judgment might be biased and debatable. Humans are prone to fallacies, and critical self reflection is the foundation of science. By considering the concept of practical relevance of a formal level, a mathematically and logically correct evaluation of the practical relevance of an observed effect or of some research results is possible.

Just because decision theoretic concepts were employed in the definitions provided within this paper does not mean that these concepts are dispensable in the absence of a formal representation of the notion of practical relevance. As outlined (Section~\ref{sec:effect_context}), elaborations in the context of practical relevance did indeed -- at least implicitly -- employ decision theoretic concepts, yet in an informal way. Accordingly, the elaborations within this paper aim only to set out these implicit decision theoretic considerations.

While it is easier to see that there is an underlying decision problem in an applied scientific investigation, it might not be that apparent in the context of foundational scientific work. Typically, an attempt to describe the statistical analysis of foundational investigations in terms of statistical decision theory yields loss functions that appear to be quite artificial, employing default loss values \citep[see e.g.][]{Berger1985}. Potential actions might be rather unspecific, such as e.g.\ ``dismiss hypothesis $H_0$'', ``do not dismiss hypothesis $H_0$'', or ``follow the line of research that is in accordance with hypothesis $H_0$''. In such a context, it might be argued that it is beneficial to refrain from making decisions \citep[e.g.][p.~110]{Rouder2018}, putting an emphasis on describing the observed data, allowing others to use it in their specific context. Also, there might be scientific research situations in which a sharp null hypothesis might be reasonable \citep[see e.g.][]{Heck2020}. If there are good reasons to do so, a sharp null hypothesis, that might not relate to a practical context, might naturally be employed. Science is very versatile and no single method or consideration does apply to every scientific context. Yet, for all those scientific investigations that are actually interested in the practical implications of their results, the definitions given within this paper might come into play.

Naturally, the question arises, whether it is necessary to have a fully specified loss function $L$ to assess the practical relevance of an effect or of a result. The unambiguous specification of such a loss function might be seen as an unsolvable task within an applied context. Information about the consequences of respective actions is expected to be scarce, vague, ambiguous, and partial, yet it should be condensed into a single quantitative entity. To make things worse, this should be done for all possible parameter values $\theta$ (of which there are frequently infinitely many) and for all actions. Consequently, it appears that the willingness of applied scientists to employ a decision theoretic account goes down to the necessity of having a loss function.

Of course, if such a loss function $L$ is fully available, then a decision theoretic analysis can be performed \citep[as e.g.\ outlined in][]{Berger1985,Robert2007}. However, as can be seen in the definitions within this paper, not all information from a loss function is required to determine the practical relevance of an effect or to specify hypotheses such that they incorporate the notion of practical relevance completely. In that, it is possible to assess the practical relevance without fully knowing the loss function. Instead, the researcher has to merely gather all available information about the consequences of the respective actions and separate the parameter space according to the preference for each action. In detail, the procedure is as follows:
\begin{itemize}
\item Think about what your research should be used for.
\item Explicitly state and describe the actions $a_0$ and $a_1$ of the decision problem that represents the purpose of your research.
\item After specifying the (parametric) statistical model, gather all available information about the consequences of deciding for $a_0$ and $a_1$ in dependence of the parameter value.
\item Using this information, determine for each parameter value which of both actions should be preferred over the other.
\end{itemize}
This leads to two parameter sets that define hypotheses that completely incorporate the notion of practical relevance. In that, hypotheses were specified reasonably w.r.t.\ the practical purpose of the investigation.

In general, this is considered to be an applied, not a statistical, problem: It is the applied scientists who have the knowledge about the practical context of interest, such that it is them who can best specify the hypotheses reasonably \citep[see e.g.][]{Kirk1996, Kirk2001, Morey2011, Lakens2017, Lakens2018, Kruschke2018}. Usually, the statisticians, who develop a statistical method, do not know the specific research context. Even further, any recommendation about default hypothesis specifications cannot match with all the different contexts a statistical method can be applied in. In many scientific fields, comprehensive guidelines on how to specify hypotheses reasonably are the exception rather than the rule. In that sense, the procedure above is a general guideline that might help applied scientists to specify their hypotheses reasonably, without being restricted to the characteristics of a specific field of applied science.

In the field of methodologies, there are plenty of methods that require statistical hypotheses to be specified reasonably w.r.t.\ the underlying practical purpose. Examples are equivalence tests \citep[see e.g.][]{Lakens2017, Lakens2018} in the frequentist setting, and Bayes factors \citep[see e.g.][]{Morey2011,Hoijtink2019, Heck2020} or decision rules that consider the region of practical equivalence (ROPE) around a null value \citep[see e.g.][]{Kruschke2015, Kruschke2018} in the Bayesian setting. The present elaboration might be helpful for these methodologies. Yet, it should be noted that the mentioned methodologies, which employ reasonably specified hypotheses, do not yield an optimal action in the context of a practical decision problem as their result: Equivalence tests result in classic tests decisions about rejecting or not rejecting a hypothesis, Bayes factors quantify evidence, and ROPE-based decision rules accept or reject a parameter null value (or withdraw the decision). In order to find the optimal action in the context of a practical decision, further loss considerations with regard to this practical decision need to be performed.

Accordingly, another alternative is to use a hypothesis-based decision theoretic analysis. Then, however, it is necessary to provide  additional quantitative specifications in the context of the loss function, than merely specifying the hypotheses reasonably. Yet, it appears that -- in the context of hypothesis-based decision theory with only two hypotheses and two actions -- it is only one loss value that needs to be specified, relating the consequences of the type-I-error (decide for $a_1$ if $H_0$ is true) to the consequences of the type-II-error (decide for $a_0$ if $H_1$ is true). Further, this loss value might also be allowed to be interval-valued \citep[in this regard, cp.][]{Walley1991, Augustin2014}, such that a range of plausible values might be specified, representing a rather robust loss specification. In that sense, the specification requirements of the loss function are extremely mitigated, allowing its employment in applied empirical investigations. How to derive the optimal action in a Bayesian setting with such a robust loss specification is elaborated on elsewhere \citep[][]{Schwaferts2019, Schwaferts2021, Schwaferts2020}.

\section{Discussion}\label{sec:discussion}

Critics might argue that the definitions of practical relevance, as provided within this paper, merely shift the difficulty of specifying hypotheses reasonably w.r.t.\ a practical purpose to the difficulty of specifying a loss function. Yet, it needs to be distinguished between
\begin{itemize}
\item \textit{[Understanding]} formulating these definitions in an attempt to understand the notion of practical relevance on a formally exact level (which is the purpose of this paper),
\item \textit{[Development of Methodologies]} developing statistical methodologies for applied scientists (which might be motivated by the elaborations within this paper and can be located within the frameworks depicted in \citep[][]{Schwaferts2019, Schwaferts2021, Schwaferts2020}), and
\item \textit{[Promotion of Methodologies]} claiming that certain statistical methodologies are appropriate in a variety of different contexts.
\end{itemize}
It is not the formulation of these definitions that shifts the difficulty to the specification of the loss function. Formulating these definitions merely generates understanding. If there is a practical purpose, then there is an underlying decision problem. A statistical analysis that wants to derive conclusions about the practical relevance of an effect or of the results needs to consider this underlying decision problem. In this context, the difficulty of specifying a loss function has -- although mainly hidden -- always been there, and current methodologies that try to circumvent loss considerations might be suboptimal in assessing the practical relevance of the observed effect or of the obtained results. Accordingly, it is not a shift in difficulty, but a disclosure of where the difficulty truly is. As outlined (Section~\ref{sec:practice}), the direction of the development of appropriate methodologies might be indicated, but their development is another issue. Naturally, of importance for this development is how to deal with the difficulties in the specification of the loss function. Yet, ignoring loss considerations at all cannot yield results that are related to the practical purpose of the study. In that, the claim that such methodologies without loss considerations yield practically relevant results should be treated with caution. Although these methodologies might appear to be applied more easily because of their ignorance to loss considerations, this cannot be an argument to promote them for scientific investigations with a practical purpose.

\bibliographystyle{diss-style}
\bibliography{Preprint_PracticalRelevance}
\end{document}